\documentclass{ws-procs10x7}

\begin{document}

\title{Study of Scalar Mesons at BES}
\author{Liaoyuan Dong \\(for the BES Collaboration)}
\address{Iowa State University, Ames, IA 50011-3160, USA
\\E-mail: dongly@slac.stanford.edu}

\twocolumn[\maketitle\abstract{
Partial wave analyses have been performed
on BES data to study scalar mesons. There is
evidence for the $\kappa$
near the $K\pi$ threshold and the pole
position is $(760\sim840) -$i$(310\sim 420)$
MeV. The $\sigma$ peak is seen in $\omega\pi^+\pi^-$
and gives an accurate pole position,
$(541\pm39)-$i$(252\pm42)$ MeV. The
$f_0(980)$ is seen in both $\phi \pi^+\pi^-$
and $\phi K^+K^-$ data.
Parameters of the Flatt\' e formula for
$f_0(980)$ are:
$M = 965 \pm 8(sta) \pm 6(sys) $ MeV,
$g_1 = 165 \pm 10(sta) \pm 15(sys)$ MeV,
$g_2/g_1 = 4.21 \pm 0.25(sta) \pm 0.21(sys)$.
The $J/\psi\to \phi \pi^+\pi^-$ data require
$f_0(1790)\to\pi^+\pi^-$, distinct from
$f_0(1710)\to K^+K^-$. Also
$f_0(1370)$ is seen clearly in $\phi \pi^+\pi^-$ data.
}]

\section{Introduction}%1
\par
{\hskip 0.4cm}

The scalar mesons are one of the most controversial subjects
in hadron physics.
Bellow 1.9 GeV, the Particle Data Group~\cite{pdg} lists the following
$I=0$ scalar states: $f_0(600)$(or $\sigma$), $f_0(980)$, $f_0(1370)$,
$f_0(1500)$, $f_0(1710)$ and two $I=1/2$ scalar states:
$K^*_0(800)$(or $\kappa$), $K_0^*(1430)$.
Scalar mesons have been traditionally studied in scattering
experiments. However, in these experiments the mesons can be
difficult to disentangle from nonresonant background.
Thus $\sigma$ and $\kappa$ are controversial;
some $f_0(980$) and $f_0(1370)$ have poorly
determined parameters.

   Recently, based on 58 million $J/\psi$ events and 14 million $\psi(2S)$
events collected with the Beijing Spectrometer (BES II) detector~\cite{bes2},
The scalar mesons have been studied by performing partial wave analysis
on many channels of BES data.
In this paper, we present some of the results from such study at BES.

\section{Results from BES} %2

\subsection{\bf\boldmath $\kappa$ in $J/\psi\to K^+K^-\pi^+\pi^-$} %2.1
\par
{\hskip 0.4cm}

\begin{figure}[htbp]
\centerline{\includegraphics[width=6.0cm,height=5.0cm]{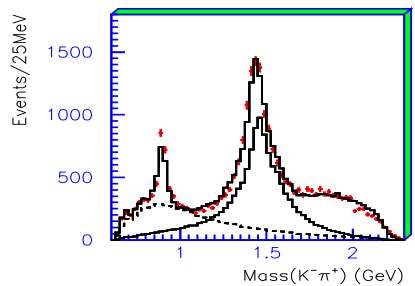}}
\caption{$K^+\pi^-$ combinations are selected in the mass range
$892 \pm 100$ MeV from $J/\psi \to K^+K^-\pi^+\pi^-$ data.
The figure shows the invariant mass distribution of accompanying
$K^- \pi^+$ pairs (crosses).
The upper full histogram shows the maximum likelihood fit,
the lower one shows the $K_0^*(1430)$ contribution,
and the dashed histogram the $\kappa$ contribution.}
\label{fig:kappa}
\end{figure}

Events over all of the 4-body phase space for
$J/\psi\to K^+K^-\pi^+\pi^-$ have been fitted.
We find evidence for the $\kappa$ in the process
$J/\psi \to K^*(890)\kappa$, $\kappa \to (K\pi)_S$.
We select a $K^+ \pi ^-$ pair in the $K^*$ mass range $892 \pm 100$ MeV;
Figure~\ref{fig:kappa} then shows the projection of the mass of the other $K^- \pi ^+$
pair.

The data are fitted with a form for the $\kappa$ containing
an Adler zero in the width. The pole position for $\kappa$ is at
$(760 \pm 20(sta) \pm 40(sys)) - $i$(420 \pm 45(sta) \pm 60(sys))$
MeV.

In Figure~\ref{fig:kappa},
the upper full histogram shows the maximum likelihood fit,
the lower full histogram shows the $K_0^*(1430)$ contribution
and the dashed histogram the $\kappa$ contribution.
There are strong destructive interference between the $\kappa$ and $K_0^*(1430)$.
The pole for $K_0^*(1430)$ lies at
$(1433 \pm 30(sta) \pm 10(sys))- $i$(181 \pm 10(sta) \pm 12(sys))$ MeV.

For the restricted region of phase space $\bar{K^*}(892)K^+\pi^-$
in $K^+K^-\pi^+\pi^-$ data, two other independent analyses have been performed.
Both favor strongly that the low mass enhancement of the
$K^+\pi^-$ system is a resonance. The $0^+$ resonances
$\kappa$ is highly necessary in both fits.
The average values of pole position for $\kappa$ is determined
to be
$(841\pm78^{+81}_{-73}) - $i$ (309 \pm 91 ^{+48}_{-72})$ MeV.

\subsection{\bf\boldmath $\sigma$ in $J/\psi\to \omega \pi^+\pi^-$} %2.2
\par
{\hskip 0.4cm}

  In $J/\psi\to \omega \pi^+\pi^-$, there are conspicuous
$\omega f_2(1270)$ and $b_1(1235)\pi$ signals.
At low $\pi \pi$ mass, a large, broad peak due to the $\sigma$ is
observed.

\begin{figure}[htbp]
\centerline{\includegraphics[width=6.0cm,height=5.0cm]{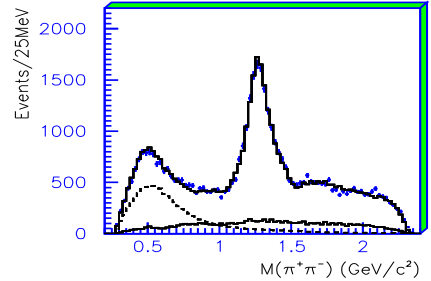}}
\caption{The $\pi^+\pi^-$ invariant mass distribution from
$J/\psi \to \omega \pi^+\pi^-$ (crosses).
The upper full histogram shows
the maximum likelihood fit, the lower full histogram
corresponds to the background estimated from
$\omega$ sidebins, and the dashed histogram
shows the $\sigma$ contribution.}
\label{fig:wpp}
\end{figure}

    Figure~\ref{fig:wpp} shows the $\pi^+\pi^-$ invariant mass distribution from
$J/\psi \to \omega \pi^+\pi^-$. 
Partial wave analyses have been performed on this channel using two
methods~\cite{sigma}. In the first method, the whole mass region
of $M_{\pi^+\pi^-}$ which recoils against the $\omega$ is analyzed,
the $\omega$ decay information is used, and the background is subtracted by
sideband estimation. For the second method, the region $M_{\pi^+
\pi^-} < 1.5$ GeV is analyzed, and the background is fitted by
$5\pi$ phase space.

   The upper full histogram in Figure~\ref{fig:wpp} shows the maximum likelihood
fit from first method,
the dashed histogram shows the $\sigma$ contribution.

  Different analysis methods and four parametrizations of
the $\sigma$ amplitude give consistent results for the $\sigma$ pole.
The average pole position is determined to be
$(541 \pm 39) - $i$(252 \pm 42))$ MeV.

  Recently, an analysis of $\psi(2S)\to\pi^+\pi^- J/\psi$ has been
performed to study the $\sigma$. The pole position of $\sigma$
is consistent with that from $J/\psi\to \omega \pi^+\pi^-$.

\subsection{\bf\boldmath Study of $J/\psi\to \omega K^+K^-$} %2.3
\par
{\hskip 0.4cm}

\begin{figure}
\centerline{\includegraphics[width=6.0cm,height=5.0cm]{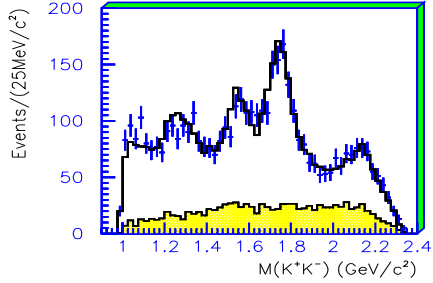}}
\caption{The $K^+K^-$ invariant mass distribution from
$J/\psi \to \omega K^+K^-$ (crosses).
The full histogram shows
the maximum likelihood fit and the shaded histogram
the background estimated from $\omega$ sidebands.}
\label{fig:wkk}
\end{figure}

   Figure~\ref{fig:wkk} shows the $K^+K^-$ invariant mass distribution from
$J/\psi \to \omega K^+K^-$. The shaded area indicates background
events from the sideband estimation.
A partial wave analysis has been
performed~\cite{wkk}, the full histogram in Figure~\ref{fig:wkk} shows
the maximum likelihood fit.

A dominant feature of $J/\psi\to \omega K^+K^-$ is $f_0(1710)$,
the present data are consistent with earlier studies which
identify $J=0$. The fitted $f_0(1710)$ optimises at
$M = 1738 \pm 30$ MeV, $\Gamma = 125 \pm 20$ MeV.

In $J/\psi \to \omega \pi ^+ \pi ^-$~\cite{sigma},
there is no definite evidence for the presence
of $f_0(1710)$ ; if its mass is scanned, there is no
optimum around 1710 MeV/c$^2$, and the fitted $f_0(1710)$
is only $0.43 \pm 0.21$\% of $\omega \pi^+\pi^-$.
In the $\omega K^+K^-$ data presented here,
the $f_0(1710)$ intensity is $(38 \pm 6)\%$ of the data within
the same acceptance as for $\omega \pi ^+ \pi ^-$.
The branching fraction for $J/\psi \to \omega f_0(1710)$,
$f_0(1710)\to K^+K^-$ is $(6.6 \pm 1.3) \times 10^{-4}$.
We find at the 95\% confidence level
\begin {equation} \frac {BR(f_0(1710) \to \pi \pi )}
{BR(f_0(1710) \to K\bar K )} < 0.11,
\end {equation}
where all charge states for decay are taken into account.

\subsection{\bf\boldmath Study of $J/\psi\to \phi \pi^+\pi^-$
                      and $J/\psi\to \phi K^+K^-$} %2.4
\par
{\hskip 0.4cm}

\begin{figure}[htbp]
\centerline{\includegraphics[width=6.0cm,height=5.0cm]{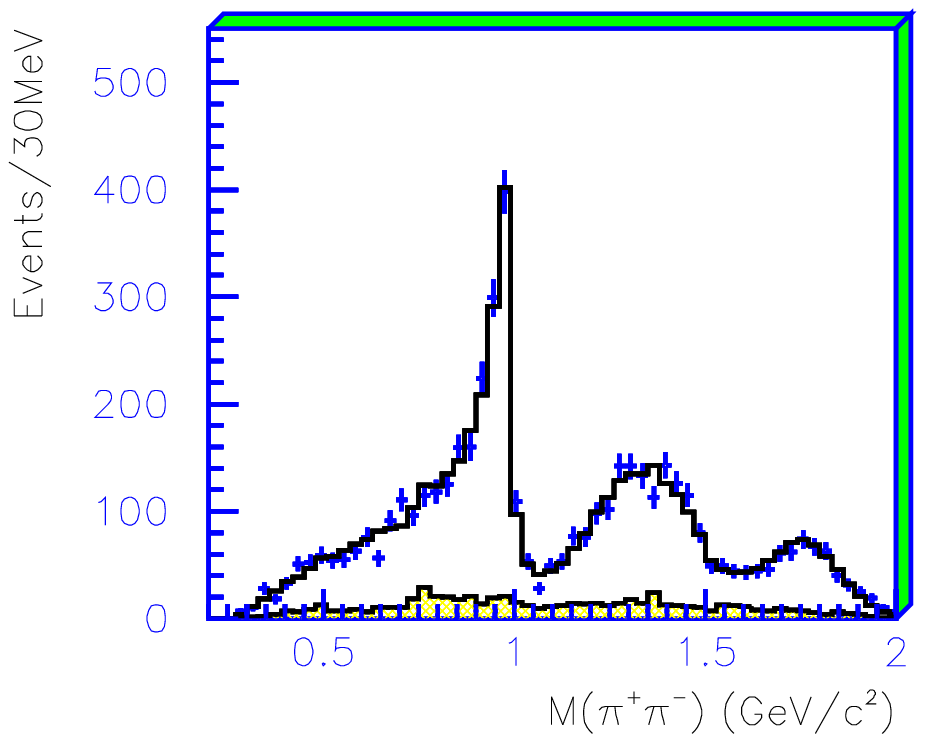}}
\caption{The $\pi^+\pi^-$ invariant mass distribution from
$J/\psi \to \phi \pi^+\pi^-$ (crosses).
The full histogram shows
the maximum likelihood fit and the shaded histogram
the background estimated from
$\phi$ sidebins.}
\label{fig:fpp}
%\end{figure}
%\begin{figure}
\centerline{\includegraphics[width=6.0cm,height=5.0cm]{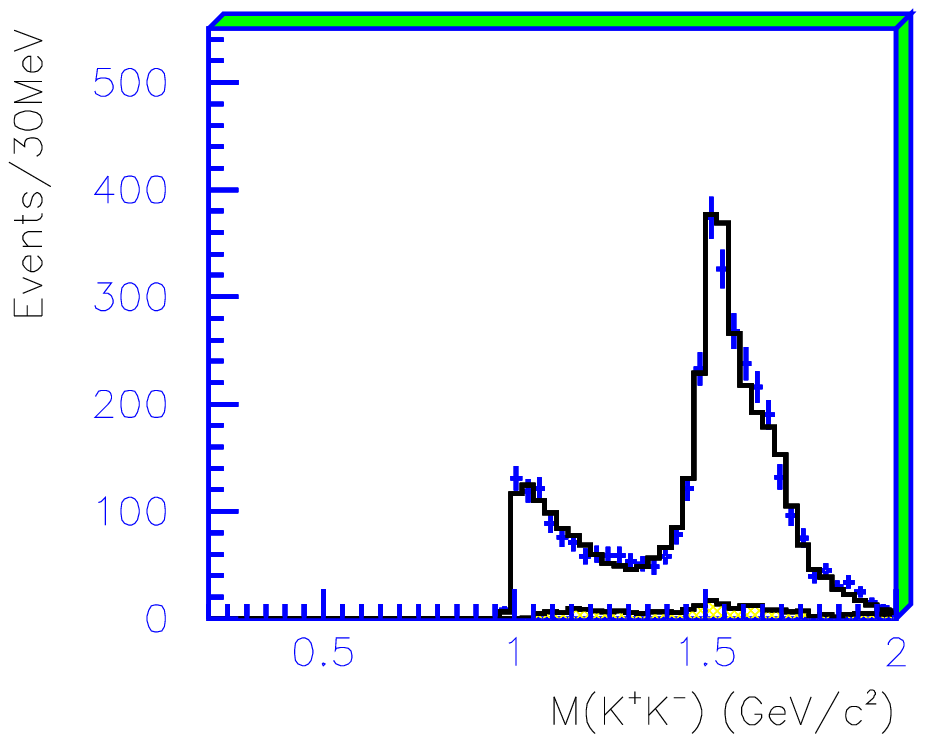}}
\caption{The $K^+K^-$ invariant mass distribution from
$J/\psi \to \phi K^+K^-$ (crosses).
The full histogram shows
the maximum likelihood fit and the shaded histogram
the background estimated from
$\phi$ sidebins.}
\label{fig:fkk}
\end{figure}

Figure~\ref{fig:fpp} shows the $\pi^+\pi^-$ invariant mass distribution from
$J/\psi \to \phi \pi^+\pi^-$. Figure~\ref{fig:fkk} shows thew $K^+K^-$ invariant mass
distribution from $J/\psi \to \phi K^+K^-$.
In Figure~\ref{fig:fpp} and~~\ref{fig:fkk}, the shaded histogram
corresponds to the the background estimated from
$\phi$ sidebins.

The $\phi \pi ^+\pi ^-$ and $\phi K^+K^-$ data are fitted
simultaneously by using partial wave analysis~\cite{fxx},
constraining resonance masses and widths to be
the same in both sets of data.
The full histogram in Figure~\ref{fig:fpp} and~~\ref{fig:fkk} show the maximum likelihood fit.

The $f_0(980)$ is observed clearly in both sets of data.
The Flatt\' e form:
\begin {equation} f=\frac {1}{M^2 - s - i(g_1\rho _{\pi \pi } +
g_2\rho _{K\bar K})}.
\end {equation}
has been used to fit the $f_0(980)$ amplitude.
Here $\rho$ is Lorentz invariant phase space, $2k/\sqrt {s}$, where $k$
refers to $\pi$ or $K$ momentum in the rest frame of the
resonance.
The present data offer the opportunity to determine the
parameters of $f_0(980)$ accurately:
$M = 965 \pm 8(sta) \pm 6(sys) $ MeV,
$g_1 = 165 \pm 10(sta) \pm 15(sys)$ MeV,
$g_2/g_1 = 4.21 \pm 0.25(sta) \pm 0.21(sys)$.

The $\phi \pi \pi$ data also exhibit a strong peak centred at
$M = 1335$ MeV.
It may be fitted with $f_2(1270)$ and a dominant $0^+$ signal
made from $f_0(1370)$ interfering with a smaller $f_0(1500)$ component.
There is definite evidence that the $f_0(1370)$ signal is
resonant, from interference with $f_2(1270)$.
The Mass and width of $f_0(1370)$ are determined to be:
$M = 1350 \pm 50$ MeV and $\Gamma = 265 \pm 40$ MeV.

There is also a definite signal from $f_0(1790) \to \pi ^+\pi ^-$
with $M = 1790 ^{+40}_{-30}$  MeV, $\Gamma = 270 ^{+60}_{-30}$ MeV.
It cannot arise from $f_0(1710)$, since the branching fraction ratio
$K\bar K/\pi\pi$ for $f_0(1790)$ is a factor 14 lower than that reported in
Ref.~4 for $f_0(1710)$. The large discrepancy in branching fractions
implies the existence of two distinct states $f_0(1790)$ and
$f_0(1710)$, the $f_0(1790)$ decaying dominantly to $\pi\pi$
and the $f_0(1710)$ dominantly to $K\bar K$.
The $f_0(1790)$ is a natural candidate for the radial excitation of
$f_0(1370)$ and behaves like $f_0(1370)$.

For $\phi K^+K^-$ data, there is a conspicuous peak due
to $f_2'(1525)$, but there is a shoulder on its upper side.
This shoulder is fitted mostly by $f_0(1710)$ interfering with
$f_0(1500)$; there is also  a
possible small contribution from $f_0(1790)$ interfering
with $f_0(1500)$.

\subsection{\bf\boldmath Study of $J/\psi\to \gamma \pi\pi$} %2.5
\par
{\hskip 0.4cm}

\begin{figure}[htbp]
\centerline{\includegraphics[width=6.0cm,height=5.0cm]{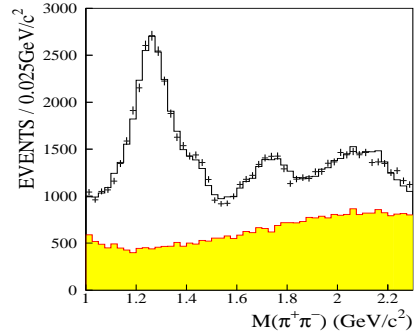}}
\caption{The $\pi^+\pi^-$ invariant mass distribution from
$J/\psi \to \gamma \pi^+\pi^-$ (crosses).
The full histogram shows
the maximum likelihood fit and the shaded histogram
corresponds to the $\pi^+\pi^-\pi^0$ background.}
\label{fig:gpp}
\end{figure}

Figure~\ref{fig:gpp} shows the $\pi^+\pi^-$ invariant mass distribution from
$J/\psi \to \gamma \pi^+\pi^-$. A partial wave analysis is carried out
in the 1.0-2.3 MeV $\pi\pi$ mass range.
There are two $0^{++}$ states in the 1.45 and 1.75 GeV
respectively. One $0^{++}$ state peaks at a mass of
$1466\pm6\pm16$ MeV with a width of $108^{+14}_{-11}\pm21$
MeV, which is approximately consistent with $f_0(1500)$.
However, due to the large interference between S-wave states, a
possibility contribution from $f_0(1370)$ cannot be excluded.

A strong production of the $f_0(1710)$ signal was observed
in the partial wave analysis of $J/\psi\to\gamma
K \bar{K}$~\cite{gkk}, with a mass of $1740\pm4^{+10}_{-25}$~MeV and a width of
$166^{+5+15}_{-8-10}$~MeV.
If the $0^{++}$ state at $\sim$ 1.75 GeV observed here
is interpreted as coming from $f_0(1710)$,
we obtain the $\pi\pi$ to $K\bar K$ branching ratio
as
$
\frac{\Gamma(f_0(1710)\to\pi\pi)}{\Gamma(f_0(1710)\to K\bar{K})}
=0.41^{+0.22}_{-0.18}.
$
This value is slightly higher than in
$\omega \pi^+\pi^-$ and $\omega K^+K^-$~\cite{wkk}.
Hence, an alternative interpretation for this $0^{++}$
state is the $f_0(1790)$.

\section{Summary}%1
\par
{\hskip 0.4cm}

Partial wave analyses have been performed
of BES data to study the scalar mesons. The $\kappa$
near the $K\pi$ threshold is needed and the pole
position is $(760\sim840) -$i$(310\sim 420)$
MeV. The $\sigma$ is seen clearly in $\omega\pi^+\pi^-$
and gives an accurate pole position,
$(541\pm39)-$i$(252\pm42)$ MeV. The
$f_0(980)$ is observed in both $\phi \pi^+\pi^-$
and $\phi K^+K^-$ data with
$M = 965 \pm 8(sta) \pm 6(sys) $ MeV,
$g_1 = 165 \pm 10(sta) \pm 15(sys)$ MeV,
$g_2/g_1 = 4.21 \pm 0.25(sta) \pm 0.21(sys)$.
The $J/\psi\to \phi \pi^+\pi^-$ data require
$f_0(1790)\to\pi^+\pi^-$, distinct from
$f_0(1710)\to K^+K^-$.
Also a peak due to $f_0(1370)$ is seen clearly
in $\phi \pi^+\pi^-$ data.

\end{document}